\documentclass[twocolumn,showpacs,preprintnumbers,amsmath,
amssymb]{revtex4}
\usepackage{graphicx}
\usepackage{dcolumn}
\usepackage{bm}

\begin{document}

\vspace{5cm}

\title{Self-similarity and singularity formation in a coupled system of Yang-Mills-dilaton evolution equations}

\author{E.E. Donets} \email{edonets@sunhe.jinr.ru, donets@msi.se
}
\author{O.I. Streltsova}
\author{T.L. Boyadjiev}
\affiliation{Joint Institute for Nuclear Research,
141980 Dubna, Russia}

\date{\today}

\begin{abstract}
We study both analytically and numerically a coupled system of
spherically symmetric SU(2) Yang-Mills-dilaton equation in 3+1
Minkowski space-time. It has been found that the system
admits a hidden scale invariance which becomes transparent if a
special ansatz for the dilaton field is used. This choice
corresponds to transition to a frame rotated in the $\ln r-t$ plane at
a definite angle. We find an infinite countable family of
self-similar solutions which can be parametrized by the $N$~-- the
number of zeros of the relevant Yang-Mills function. According to
the performed linear perturbation analysis,  the lowest solution
with $N=0$ only occurred to be stable. The Cauchy problem has been
solved numerically for a wide range of  smooth finite energy
initial data. It has been found that if the initial data exceed
some threshold, the resulting solutions in a compact region
shrinking to the origin, attain the lowest $N=0$ stable
self-similar profile, which can pretend to be a global stable
attractor in the Cauchy problem. The solutions live a finite time
in a self-similar regime and then the  unbounded growth of the
second derivative of the YM function at the origin indicates a
singularity formation, which is in agreement with the general
expectations for the supercritical systems.
\end{abstract}

\maketitle
\section{ Introduction}

Singularity formation in nonlinear evolution equations is a
general problem which arises in various branches of mathematical
physics. The term singularity means that the solutions to the
evolutional differential equations cease to be differentiated at
some point/region. However, the strong-field regime of the
approach to the singularity can provide some universality and
additional symmetries in the solutions behavior that is  of
particular interest in the corresponding theoretical field models,
including gravity.

 It is well-known the space-time singularities are the most generic
features of Einstein's equations \cite{sing} and it is believed
that the mixmaster type singularity can pretend to be a generic
one. However, the nature of the space-time singularity is
model-dependent and various matter fields provide various scenario
of the singularity formation. In the typical case of a
gravitational collapse the late-time dynamics of the space-time
singularity development is hidden under the formed event horizon.
However the last decade studies of the massless fields collapse
initiated in a pioneer work by M. Choptuik \cite{ch1} opened a new
direction for understanding  the singularity formation dynamics.

It was realized  \cite{ch1} that black holes with an arbitrary
small masses would be obtained and the mass scale low was
discovered. This was called Type-II behavior which is
characterized by a mass gap absence in the black hole spectrum.
The limiting case of a vanishing black hole mass is of particular
interest. It has been found these critical solutions observed at
the threshold of the black hole formation are discretely
self-similar.

Later on, the numerical studies of the gravitational collapse of
the self-interacting massless fields were performed \cite{ch2},
\cite{sky} and the Type-I behavior was observed as well. The
Type-I behavior is characterized by a finite mass gap in the black
hole spectrum. This Type-I behavior takes place if the considered
system of Einstein-matter equations admits static finite energy
asymptotically flat solutions. In this case the smallest black
hole has a finite mass which is equal to the mass of the lowest
static solution. In the Einstein-Yang-Mills (EYM) system of
equations the $N=1$ Bartnick-McKinnon (BK) static solution
\cite{r4} is the lowest mass static solution which determines a
minimal mass of the formed black hole. Moreover, this $N=1$ BK
solution is occurred to be an intermediate attractor which the
collapsing solution should attain in order to turn to the Type-I
black hole formation scenario \cite{ch2}, \cite{gund}.

On the other hand the analysis of the blowup in the nonlinear wave
equations in various field models without gravity showed the
singularity formation in gravity and blowup in the nonlinear wave
equations share many common features \cite{b1}, \cite{b2},
\cite{b3}, \cite{lin}. Based on these observations, P. Bizon and
Z. Tabor put forward the conjecture \cite{b1} that all basic
properties of the gravitational collapse of massless fields such
as universality, self-similarity and mass scaling, originally
observed for Einstein's equations are just the basic properties of
a wide class of a supercritical evolution PDEs. This class
includes Einstein equations, Yang-Mills equation in $5+1$
Minkovski space and many others.

Following  this  conjecture, we consider a coupled system of
Yang-Mills-dilaton (YM-dilaton) equations in $3+1$ Minkovski space
which is of interest for several reasons. First of all, this
system is a truncated version of a theoretical field model
inspired by the heterotic string. Then the dilaton field itself
which is also called a scalar graviton provides many key features
characteristic of a gravity. For example the static system of
spherically symmetric $SU(2)$ Yang-Mills-dilaton equations has a
countable infinite set of  regular finite mass solutions
\cite{r10} which are similar to the BK solutions in the
Einstein-Yang-Mills system of equations. Note that a  similar set
of the regular solutions exists in the Einstein-Yang-Mills-dilaton
system \cite{eymd1}, \cite{eymd2}. This fact stresses the dominant
role of the Yang-Mills field for the YM-dilaton, EYM and
EYM-dilaton systems dynamics. One can expect that the YM-dilaton
model should exhibit all main features relevant for the EYM and
EYM-dilaton models. Hence a study of the singularity formation in
the YM-dilaton system should shed  new light on a singularity
development hidden under the formed event horizon in the EYM and
EYM-dilaton models.

We have found that the considered system of spherically symmetric
$SU(2)$ YM-dilaton equations admits a hidden scale invariance. The
corresponding criticality index is equal to $+1$ similar to the
Einstein's equations in $3+1$ dimension and Yang-Mills equation in
$5+1$ Minkovski space-time. This means the YM-dilaton is
supercritical and singularity should arise if the initial data
exceeds some threshold. Note a pure Yang-Mills system of equations
in $3+1$ Minkovski space-time is subcritical and the solutions
should remain everywhere smooth during the evolution \cite{b21}.
We have found the late-time asymptotics of the YM-dilaton
solutions prior to the singularity formation is universal and is
described by the lowest $N=0$ self-similar profile. This $N=0$
profile for YM field is occurred to be similar to those found in
\cite{b1}, \cite{b2} for pure YM field in $5+1$ Minkovski space.
We have shown the whole family of the self-similar solutions
exists labelled by the $N=0,1,2,3\ldots$ $\infty$ -- number of the nodes
of the relevant YM function.

 The paper is organized as follows. In the next Section we
introduce main definitions and discuss the scale properties of the
obtained YM-dilaton system of spherically symmetric equations. In
the third Section the self-similar solutions of the YM-dilaton
system  and their linear stability are considered. And in the
fourth Section the results of the numerical simulations of the
evolutional Cauchy problem are discussed. We conclude with the
main results and discuss briefly some open questions in the last
Section.

\section{ Main equations and definitions}

We consider a coupled system of Yang-Mills-dilaton fields, which is given by the action:
\begin{equation} \label{action}
    S = \frac{1}{4\pi}\int\left[\frac{1}{2}(\partial\Phi)^2- \frac{\exp(k\Phi)}{4g^2}
     F^{a\mu\nu}F^{a}_{\mu\nu}\right]d^3x\,dt\,,
\end{equation}
\noindent
where $\Phi$ --- dilaton field and $F^{a\mu\nu}$ --- Yang-Mills field. Note that this action is a truncated version of the bosonic part of the heterotic string effective action in four
dimensions in Einstein frame \cite{l18}.

In the case of the spherical symmetry, the dilaton field and the
SU(2) Yang-Mills potential can be parameterized in terms of two
independent functions
 $\Phi(t,r)$ and $f(t,r)$ as follows:
\begin{equation}
    \Phi=\Phi(r,t),\quad A^a_t=0\,,
\quad A^a_i= \epsilon_{aik}\frac{x^k}{r^2}\,
     \left(f(r,t)-1 \right).
\end{equation}

After substitution of this ansatz into the action and to the field
equations, the rescaling $\Phi\to\Phi/k,\,
r\to (k/g)r,\, t\to (k/g)t$ and $S\to g^2 S$
 removes the dependence on $k$ and $g$ and therefore we can put
$k=1$, $g=1$
in what follows without restrictions. After integrating out the
angular variables, the resulting action and the field equations
become:
\begin{eqnarray}
S =& - & \int \left[\frac{1}{2} r^2 {\Phi\,'}^2 - \frac{1}{2} r^2
{\dot\Phi}^2 \right. \nonumber \\ [0.1cm]  &+&\left.
e^{\Phi}\left({f\,'}^2-{\dot
f}^2+\frac{(f^2-1)^2}{2r^2}\right)\right]\,dr\,dt,
\end{eqnarray}
\begin{eqnarray} \label{5a}
&&\ddot f + \dot f\dot\Phi-f\,''-f\,'\Phi\,'=
\frac{f(1-f^2)}{r^2},\\[0.1cm] \label{5b}
&&\ddot\Phi-\Phi''-\frac{2\Phi'}{r}=
-\frac{e^\Phi}{r^2}\left(f\,'^2-\dot
f^2 + \frac{(f^2-1)^2}{2r^2} \right).
\end{eqnarray}

Here and below (except Eq.~(\ref{vv})) dot stands for a partial
derivative with respect to $t$, while prime is a partial
derivative with respect to $r$.

It is also useful to write down the expression for the total
energy
\begin{eqnarray}
E& =&\int\limits_0^{+\infty}\left[\frac{1}{2}r^2{\Phi\,'}^2 +
\frac{1}{2} r^2 {\dot\Phi}^2 \right. \nonumber\\ [0.1cm] &+&\left.
e^{\Phi}\left({f\,'}^2+{\dot
f}^2+\frac{(f^2-1)^2}{2r^2}\right)\right]\,dr,
\end{eqnarray}
which is, of course, conserved  {\it on shell}.

For further analysis it is important to note that the system
Eqs.~(\ref{5a}, \ref{5b}) admits a hidden scale-invariant form.
Indeed, the Eq.~(\ref{5a}) is scale-invariant in sense that if
$f(t,r), \Phi(t,r)$ is a solution to equation (\ref{5a}), then
$\widetilde{f}(t,r)= f(\frac{t}{\lambda}, \frac{r}{\lambda}),
\widetilde{\Phi}(t,r)=\Phi(\frac{t}{\lambda}, \frac{r}{\lambda})$
is also a solution. The same is not true for Eq.~(\ref{5b})
because of the factor $e^{\Phi}/r^2$ on the right-hand side of
equation (\ref{5b}), which breaks the scale invariance. However,
it is possible to extract a scale invariant part $\phi(t,r)$ from
the dilaton function $\Phi(t,r)$ that makes transparent the hidden
scale invariance of the system. Indeed, if we consider the
following ansatz
\begin{equation}
\label{prphi}
    \Phi(r,t) = \phi(r,t) + 2 \ln r,
\end{equation}
the system Eqs.~(\ref{5a}, \ref{5b}), rewritten in terms of
$f(t,r)$, $\phi(t,r)$ becomes scale-invariant. The energy,
expressed in terms of the functions $f(t,r)$, $\phi(t,r)$ becomes
\begin{widetext}
    \begin{eqnarray}
        E = \int\limits_0^{\infty} dr \left\{\frac{1}{2}r^2 \phi\,'^2 + 2r \phi\,' + 2 + \frac{1}{2}r^2\dot \phi^2 +r^2 e^{\phi} \left[ f\,'^2+\dot
f^2+\frac{(f^2-1)^2}{2r^2}\right] \right\},  \label{Etot1}
     \end{eqnarray}
\end{widetext}
providing the corresponding homogeneous scale low for the energy as $E(f(\frac{t}{\lambda}), \phi(\frac{r}{\lambda}))= \lambda E(f(t,r), \phi(t,r))$. According to the PDE general theory \cite{b22} the degree $\alpha$ of the scale parameter $\lambda$ as it enters in the energy homogeneous scale low defines the criticality class of the PDE system. Criticality class of the system indicates the possibility of a singularity formation in the corresponding well-posed Cauchy problem as follows. If $\alpha < 0$ the system is subcritical, then all initially regular solutions should remain globally regular during the evolution. If $\alpha > 0$ the system is supercritical and one should expect singularity formations for a finite time for all initial data, if its exceed a some threshold values. If $\alpha = 0$, the system is critical and there are no  definite expectations on singularity formations. Since in our case we have $\alpha= +1$, the system is supercritical and one should expect singularity formations, which
will be confirmed in the last section of this paper.

The revealed invariance of the solutions under the scale dilations
$f(t,r) \to f(\frac{t}{\lambda}, \frac{r}{\lambda}), \phi(t,r) \to
\phi(\frac{t}{\lambda}, \frac{r}{\lambda})$ means that both, $t$
and $r$ variables enter effectively the solutions only in
particular combinations - $r/t$, or $t/r$. Because of the time
translation invariance it is useful to introduce some positive
constant $T$, which transforms similarly under the dilations, and
search for general scale-invariant solutions in  a self-similar
form:
\begin{equation}\label{11}
  f(r,t)= f(x),\quad \phi(r,t)= \phi(x), \quad x= \frac{T-t}{r}.
\end{equation}

The constant $T$ has the meaning of a blowup time --- absolute
value of the time in the evolution Cauchy problem, when the
expected singularity starts  development and the solution cease to
be smooth. We also will use alternatively the inverse independent
variable $\eta= \frac{1}{x} = \frac{r}{T-t}$, which is more
natural for study of the Cauchy problem. The coordinate $\eta$
covers a half of the complete Minkowski space-time only,
corresponding to the past region of the blowup point $t=T, r=0$.
The coordinate $x$ covers complete Minkowski space and we will use
it mainly for the analysis of the self-similar solutions in the
next section.

\section{ Self-similar solutions and their linear stability analysis}

In order to bring the system of the scale invariant equations to
the form, suitable for further analysis, it is convenient to
introduce a new function $s(x)$ as follows:
\begin{equation*}
    \phi(x)= \ln\left(x^2 s(x)\right).
\end{equation*}

As we will see below, $s(x)$ represents the regular part of the
function $\phi(x)$ on the semiaxis $x \in [1, \infty)$.

In terms of the functions $f(x)$ and $s(x)$ the system of PDEs (\ref{5a}, \ref{5b}) transforms to a system of ODEs:
\begin{eqnarray} \label{ssx1}
    &&f_{,x,x}+\frac{2 f_{,x}}{x}+\frac{f_{,x} s_{,x}}{s} =
\frac{f\,(1-f^2)}{1-x^2}\,,
\end{eqnarray}
\begin{eqnarray} \nonumber
\frac{s_{,x,x}}{s}-\frac{s_{,x}^2}{s^2}-x^2\,s\,f_{,x}^2-
\frac{2}{x^2}&=& \frac{2}{1-x^2}-\\[0.1cm]\label{ssx2}&
&\frac{x^2\, s \,(1-f^2)^2}{2(1-x^2)}.
\end{eqnarray}

This system has 4 singular points $x=-\infty$, $-1$, $+1$, $+\infty$.
At the first step we restrict ourselves to the interval $x \in [1,+\infty)$ which covers the interior of  the past light cone of the point $t=T, r=0$.

The natural requirement of the regularity for the YM function $f(x)$ on the past light cone of the point $t=T, r=0$ provides the following local solution of the system (\ref{ssx1}, \ref{ssx2})
near $x=1$:
\begin{eqnarray}
&&f(x)_{x \to 1}=f_1(x-1)-\frac{f_1}{8}(10+s_1)(x-1)^2 +  \nonumber\\
&&+f_1\left(\frac{31}{48}-\frac{2}{3} f_1^2+\frac{1}{96} s_1^2+
   \frac{5}{32} s_1\right)(x-1)^3+O((x-1)^4), \nonumber \\[0.1cm]
&&s(x)_{x \to 1}=4+s_1 (x-1) +     \label{ser1} \\
&&+\left(8+8 f_1^2+\frac{1}{8}s_1^2+ \frac{1}{2}s_1\right)(x-1)^2+ \nonumber \\
&&+\left(-8f_1^2+\frac{4}{3}s_1f_1^2+\frac{7}{3}s_1+\frac{1}{96}s_1^3+
\frac{7}{48}s_1^2-\frac{4}{3}\right) (x-1)^3 \nonumber \\
&&+O((x-1)^4),\nonumber
\end{eqnarray}
where $f_1$ and $s_1$ are free parameters. The regularity
requirement of the YM function $f$ at the origin on each slice
$t<T$ (until the blowup time $t=T$) leads to the following series
expansion of the solutions near $x=+\infty$, written by making use
of the inverse $\eta=1/x$ self-similar variable:
\begin{eqnarray} \label{ser2}
  f(\eta)_{\eta \to 0}&=&\pm 1+f_2 \eta^2 + \nonumber\\[0.1cm]
 &+&\frac{1}{10}f_2\left(-4s_0f_2^2+3f_2 +
\frac{10}{3}\right)\eta^4+O(\eta^6),\nonumber\\[0.1cm]
s(\eta)_{\eta \to 0}&=&s_0+s_0(s_0f_2^2-\frac{1}{3})\eta^2 + \\[0.1cm]
&+&
\frac{1}{20}s_0\left(8s_0^2f_2^4+8s_0f_2^3-\frac{8}{9}\right)\eta^4+
O(\eta^6).\nonumber
\end{eqnarray}
Here $f_2$ and $s_0$ are also free parameters.

These local solutions at the singular points satisfy appropriate
boundary conditions:
\begin{eqnarray}
  f(1)= 0\,, \quad  s(1)= 4\,,\quad f(\infty)= \pm 1\,,\nonumber\\
  f,_x(x)\mathop \to \limits_{x \to \infty} 0\,, \quad s,_x(x)\mathop
  \to \limits_{x \to \infty } 0\,. \label{gr}
\end{eqnarray}
\begin{figure}
\includegraphics[width=8.6cm]{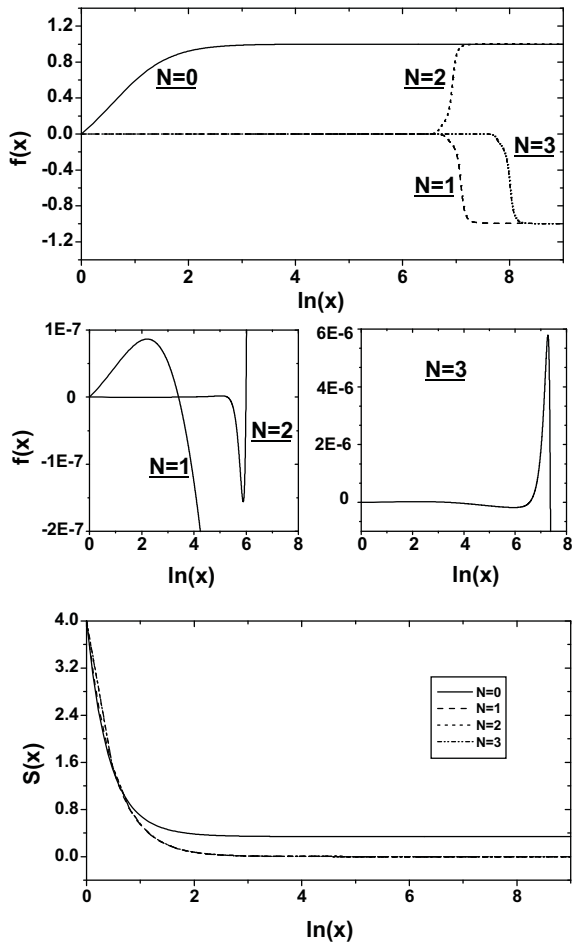}
\caption{\label{fig:1} Self-similar solutions for $N=0,1,2,3$: YM
functions $f_{N}(x)$~ -- top, dilaton functions $s_{N}(x)$~- bottom.
Note, as $N$ increases the solutions $s_{N}(x)$ tends uniformly to
the limiting curve $\frac{4}{x^2}$.}
\end{figure}
As a result we get a suitable desingularised boundary-value
problem (BVP) for the system (\ref{ssx1}, \ref{ssx2}) with the
boundary conditions of Eq.~(\ref{gr}). Alternative strategy
consists in solving of the system (\ref{ssx1}, \ref{ssx2}) as an
initial value problem starting with the series expansion
(\ref{ser1}) near the $x= 1$. In this case the problem transforms
to the shooting method search for some discrete sets of free
parameters $f_1, s_1$, so that they provide the meeting of the
local solutions to Eq.~(\ref{ser2}) at infinity. The method of
shooting technique also gives a powerful tool to prove  global
existence, developed by Breitenlohner, Forgach and Maison
\cite{BFM}. We are planing to consider the proof of the existence
in a separate paper.

In order to find the above solutions numerically, we consider the
above BVP using the method, described in details in \cite{btfy02}.
In accordance with the general expectations, we have found that
the BVP (\ref{ssx1}), (\ref{ssx2}), (\ref{gr}) admits infinitely
many solutions and all of then can be labeled by the $N$~ -- the
total number of zeros of the YM function $f(x)$ on the open
semiaxis $x \in (1, \infty)$. The solutions with the lowest $N$,
$N= 0,1,2,3$ are given in Fig.~\ref{fig:1}. One can also extract
the values of the corresponding free parameters $f_1$, $s_1$, or
$f_2$, $s_0$ as they enter the local series expansions
Eq.~(\ref{ser1}) and Eq.~(\ref{ser2}). The values of $f_1$ and
$s_1$ are displayed in Table~\ref{tab:1}.

\begin{table}[ht]
    \caption{\label{tab:1}
    The values of free parameters $f_1$,
$s_1$ for the self-similar solutions with various $N$}
      \begin{center}
        \begin{ruledtabular}
            \begin{tabular}{lcc}
                $N$& $f_1$ & $s_1$\\
                \hline
                 0       & 0.498934096775465 & -8.92179247 \\
                 1       & $1.13571 10^{-12}$ & -8.00124    \\
                 2       & $8.98718 10^{-13}$ & -8.00095    \\
                 3       & $8.06871 10^{-13}$ & -8.00068    \\
                \ldots   & \ldots            & \ldots      \\
                $\infty$ & 0.0               & -8.0        \\
            \end{tabular}
        \end{ruledtabular}
    \end{center}
\end{table}

Note that the solution which corresponds to $N= \infty$ is simply
the trivial solution $f(x)= 0$, $s(x)= 4/x^2$, that in terms of
the primary function reads $\phi(x)= \ln4$.

If $N$ is finite and $N>>1$, the solution of the system of
Eqs.~(\ref{ssx1}, \ref{ssx2}) can be easily found by applying the
linear perturbations theory around the trivial solution in the
oscillation region. Indeed, after the substitution of $f(x)=
0+\epsilon\widetilde{f}(x)$, $\phi(x) = \ln4 + \epsilon\widetilde{\phi}(x)$ into the system Eqs.~(\ref{ssx1},
\ref{ssx2}), the linearized system for the perturbations
$\widetilde{f}(x), \widetilde{\phi}(x)$ which holds at the
interval $1\ll x_0\leq x\leq x_N$ ($x_N\sim\exp(2\pi/\sqrt{3}N)$),
becomes
\begin{eqnarray}
    \frac{d^2\widetilde{f}}{dx^2}=
-\frac{1}{x^2}\widetilde{f}\,,\quad\quad
    \frac{d^2\widetilde{\phi}}{dx^2}= \frac{2}{x^2}\widetilde{\phi}\,.
\end{eqnarray}

The solution, bounded at infinity can be found in terms of elementary
functions as follows:

\begin{eqnarray}
    \widetilde{f}(x) = C_1\sqrt{x}\;{\rm \sin}\left(
\frac{\sqrt{3}}{2}\ln x+
    \delta\right)\,, \quad \widetilde{\phi} = \frac{C_2}{x}\,,
\end{eqnarray}
where $C_1$, $C_2$ and $\delta$ are some integration constants.
This oscillating solution gives a strong argument in favor of the
existence of solutions with arbitrary $N>0$ zeros of YM function.

We found the solutions, which are defined on the interval $x \in
[1, +\infty)$. Now one can analytically continue them in terms of

the functions $f(x), \phi(x)$ to the left of the point $x=1$. It
has been done numerically. The continued  functions  behave
monotonically at the interval $x \in [0,1]$ and they meet their
corresponding values at the regular point $x=0$: $f(x=0)=
f_{\infty}$, $\phi(x=0)= \phi_{\infty}$ which are also their
asymptotic values at the spatial infinity $r \to +\infty$ on each
slice $t= const$, $t<T$ in the self-similar regime. This can be
seen from the series expansion of the solutions in terms of the
$\eta$ variable at $\eta \to +\infty$ ($f_\infty,  \phi_\infty,
\bar{f_1}, \bar{\phi_1}$ are free parameters):
\begin{eqnarray}
&&f(\eta)_{\eta\to\infty}= f_\infty+{\bar f_1}\eta^{-1} +
 \nonumber\\[0.1cm]
&&+\frac{1}{2} \left(-{\bar f_1} {\bar \phi_1} + f_\infty -
f_\infty^3\right) \eta^{-2} + O(\eta^{-3}),\nonumber\\[0.1cm]
&&\phi(\eta)_{\eta\to\infty}=\phi_\infty+{\bar \phi_1}\eta^{-1} +
\\[0.1cm]
&&+\frac{1}{2}\left[2+e^{\phi_\infty}\left({\bar f_1}^2 - \frac{1}{2} +
f_\infty^2 - \frac{1}{2} f_\infty^4\right)\right] \eta^{-2} +
O(\eta^{-3}).\nonumber
\end{eqnarray}

For example, the values $f(x=0)=f_{\infty}=-0.5072593 \ldots$,
$\phi(x=0) = \phi_{\infty} = 2.1214115\ldots$ correspond to the $N=0$
solution (see below Fig.~\ref{fig:5}).

Further continuation of the solutions to the region of the
negative $x$ meets an obstacle at the next singular point $x=-1$.
It was found that all self-similar solutions (except the trivial
one), labeled by the $N=0,1,2,3, \ldots \infty$ ceased to be differentiable in the vicinity of the singular point $x=-1$ and therefore they cannot be continued smoothly to the complete manifold, covered by the $x$ coordinate.

The stability analysis of the obtained self-similar solutions can
be done in terms of the linear perturbation theory. The negative
energy states in the spectrum of the linearized system  are
usually related to the instability of the background solutions.

Now we consider the problem on the semiaxis $x \in [1, +\infty)$.
Let us introduce in addition to the variable $x$ a second
independent variable $\tau$ so that the set of the lines $x =
const$ is orthogonal to the set of the lines $\tau=const$:
\begin{eqnarray}
    x=\frac{T-t}{r},\quad \tau=-\ln\sqrt{(T-t)^2-r^2}\,.
\end{eqnarray}
The action (\ref{action}) in the $x, \tau$ variables after the
substitution of the anzats (\ref{prphi}) for the dilaton field
becomes:
\begin{widetext}
\begin{eqnarray} \label{actxtau}
    \nonumber S=&-&\int dx\, d\tau\, e^{-2\tau}\,\left\{
\frac{1}{2}\left( \frac{\partial \phi }{\partial x}\right) ^2 -
\frac{1}{2(x^2-1)^2}\left( \frac{\partial \phi}{\partial
\tau}\right) ^2+\frac{2}{(x^2-1)^2}\left( \frac{\partial
\phi}{\partial \tau}\right)-\frac{2x}{x^2-1}\left( \frac{\partial
\phi}{\partial x}\right)+ \frac{2}{x^2-1}\right.\\[0.1cm]
&+&\left. e^\phi \left[ \left( \frac{\partial \phi}{\partial
x}\right) ^2 - \frac{1}{(x^2-1)^2}\left( \frac{\partial
\phi}{\partial \tau}\right) ^2 + \frac{(f^2-1)^2}{2(x^2-1)}
\right] \right\}.
\end{eqnarray}
\end{widetext}

We look for spherically symmetric perturbed solutions of the
following form:
\begin{widetext}
\begin{eqnarray}
    f(x,\tau)=f_N(x)+\epsilon \, \sqrt{2(x^2-1)}\, v(x)e^{i\omega\tau},\quad\quad
 \phi(x,\tau)=\phi_N(x)+\epsilon\, e^{-\phi_0(x)/2}\sqrt{x^2-1}\,
 u(x)e^{i\omega\tau},
\end{eqnarray}
\end{widetext}
where the background $f_N(x),\phi_N(x)$ are some solutions of the
BVP (\ref{ssx1}),
 (\ref{ssx2}), (\ref{gr}), parametrized by the $N$~- the number
of zeros of the  YM function $f(x)$.
\begin{figure}
\includegraphics[width=8.cm]{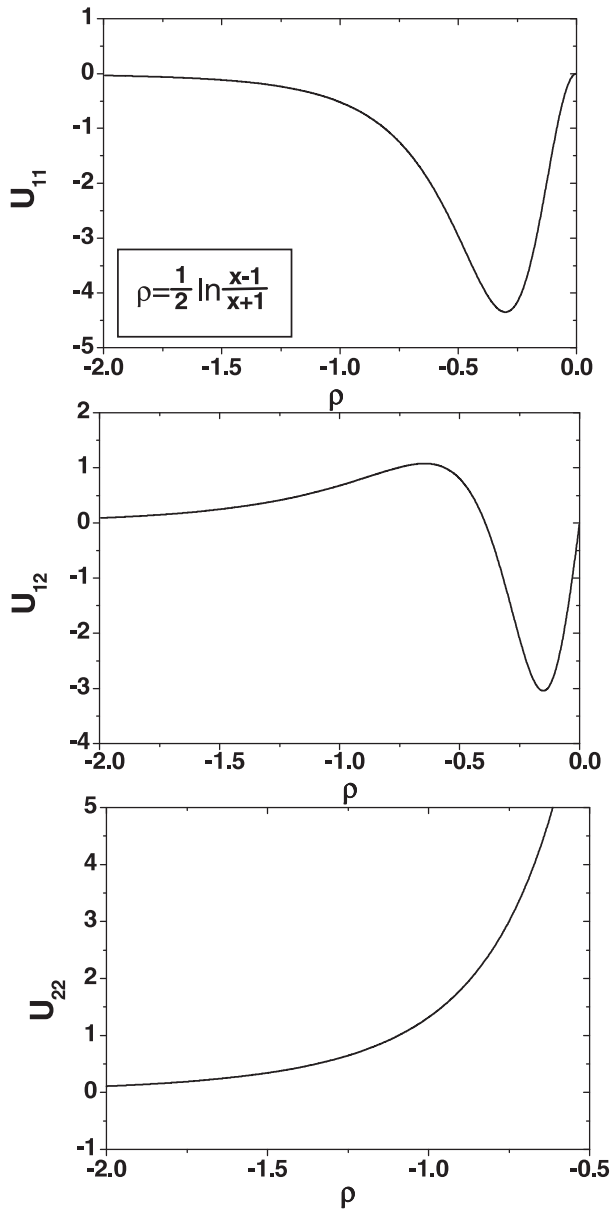}
\caption{\label{fig:2} Matrix elements $U_{ij}(\rho)$ for the
background self-similar solution with $N=0$: $U_{11}$~ -- top,
$U_{12}=U_{21}$~- middle, $U_{22}$~- bottom}.
\end{figure}
It is helpful to use the variable $\rho$ instead of the $x$:
\begin{equation}
    \rho=\frac{1}{2}\ln\left(\frac{x-1}{x+1} \right),
\quad x=\frac{1+e^{2\rho}}{1-e^{2\rho}}.
\end{equation}
Then, following the lines of the Ref.\cite{r10}, one can bring the
effective action for the perturbations to the standard form of
two-channel $\text{Schr}{\ddot o}\text{dinger}$ equation
\begin{equation}
    {\widetilde S}=-\int d\tau\, d\rho\, e^{2\tau}\,\left[
(\chi)_{,\rho}^{+}(\chi)_{,\rho}+\chi^+U\chi-(\omega^2-1)\chi^+
\chi \right],
\end{equation}
where the column $\chi$ and the
function $A(\rho)$ are introduced as follows ($\sigma_2$ is Pauli
matrix):
\begin{eqnarray}\label{xi}\nonumber
  &&\chi=e^{-i\sigma_2 A(\rho)}{v(\rho) \choose  u(\rho)},\\[0.1cm]
 &&A(\rho)=\int_0^{\rho}\frac{\exp({\phi_N(\rho)/2})f_N(\rho)_{,\rho}}
 {\sqrt{2}}\,d\rho.
\end{eqnarray}

Finally we get a matrix equation, which describes the spectrum of
the linear perturbations
\begin{equation}\label{shred}
  -\chi_{,\rho,\rho}+U \chi=\Omega^2 \chi,  \quad \Omega^2=\omega^2-1\,,
\end{equation}
where
\begin{equation}\label{matrix}
    U={U_{11}\quad U_{12} \choose  U_{21}\quad U_{22}}=
e^{-i\sigma_2A}V\,= e^{i\sigma_2 A}\,,
\end{equation}
and the elements of the matrix $V$ are
\begin{eqnarray}\label{vv}
\nonumber V_{11}&=&e^{\varphi_N}\left({f\,'_N}^2+\frac{(f_N^2-1)^2}
{2\sinh^2\rho}\right), \\
    V_{12}&=& V_{21}= \frac{1}{\sqrt{2}\sinh^2\rho}\left(\sinh^2 \rho\,
e^{\varphi_N/2}f\,'_N \right)' \\
\nonumber
    V_{22}&=&\frac{1}{2}\varphi\,''_N+ \frac{1}{4}{\varphi\,'_N}^2
+(\coth\rho )\varphi\,'_N + \frac{3f_N^2-1} {\sinh^2\rho}.
\end{eqnarray}

Note that prime in formula (\ref{vv}) stands for derivative with
respect to $\rho$ variable.

For the background solution $f_0(\rho)$, $\phi_0(\rho)$, which corresponds to the $N=0$, the matrix elements $U_{11}(\rho)$, $U_{12}(\rho)$, $U_{22}(\rho)$ are shown in Fig.~\ref{fig:2}.
Background solutions with other $N>0$ provide the matrix elements $U_{ij}$ of a similar shape.

Using the method of the phase functions shift, introduced by F.
Calogero \cite{calogero} and developed by A. Degasperis
\cite{l20}, we have found that the self-similar solutions
$f_N(\rho), \phi_N(\rho)$ with $N$ zeros of YM function $f$ have
$N$ unstable modes in the spectrum of the linear perturbations.
Hence the only self-similar solution with $N=0$ is linearly
stable.

\section{The Cauchy problem}

In this section we consider the Cauchy problem for the system of
equations (\ref{5a}, \ref{5b}) starting with regular initial data.
The goal is to study the behavior of the solutions near the
origin, their possible attaining to some self-similar solutions
and their late-time asymptotics prior to the expected blowup.

The system of equations (\ref{5a}, \ref{5b}) is a coupled system
of nonlinear wave equations. The boundary conditions at the origin
and at the radial infinity as well as the initial conditions on
the slice $t=0$ have to be defined in order to get a well-posed
Cauchy problem. In fact, such a problem with the initial and
boundary conditions imposed is called a mixed type Cauchy problem.
However, we will call it Cauchy problem, for short.

The system of equations (\ref{5a}, \ref{5b}), similarly to its
static version, have two singular  boundary points $r=0$ and
$r=+\infty$. The regularity  requirement provides the only
possible series expansion near the origin:
\begin{eqnarray}
&&f(t,r)_{r \to 0}=1-b(t)r^2+O(r^4),\nonumber\\[0.1cm]
&&\Phi(t,r)_{r \to 0}=\Phi_0(t)+\Phi_2(t)r^2+O(r^4).
\end{eqnarray}
\noindent where $b(t)$, $\Phi_0(t)$, $\Phi_2(t)$ are finite smooth
functions. Hence the corresponding boundary conditions at the
origin, which must hold during the evolution, are as follows:
\begin{eqnarray}
    \nonumber &&f(t,r=0)=1,\quad  f\,'(t,r=0)=0,\\[0.1cm]
&&\Phi(t,r=0)=\Phi_0(t), \quad \Phi\,'(t,r=0)=0.\label{boundary1}
\end{eqnarray}
\begin{figure}
\includegraphics[width=8.cm]{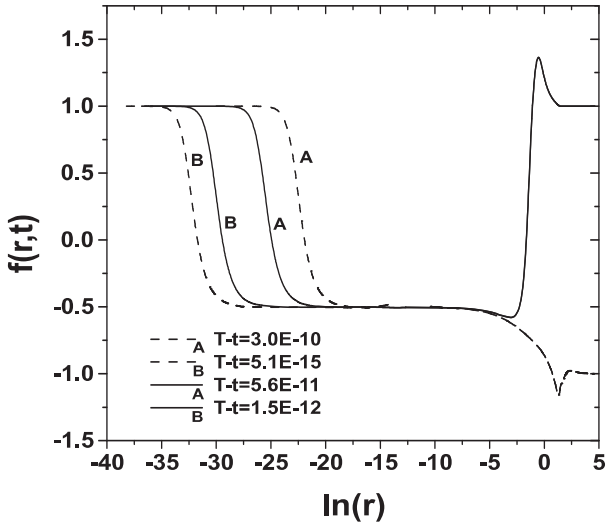}
\caption{\label{fig:3} The set of YM profiles $f(t,r)$ at the
various times prior to the singularity formation. The ingoing
waves at the small radial scales exhibit universal self-similar
behavior whereas the outgoing waves at the right side of the
figure are almost frozen at the given time scale. The solutions
correspond to the following initial data data:\newline I.
$f(0,r)=(1-a r^2)/(1+a r^2)$ -- kink-type, $a=0.281$ (dashed);
\newline II. $f(0,r)=1-Ar^2\exp(-\sigma(r-r_0)^2)$ -
Gauss-type,\newline $A=0.2, \sigma=10, r_0=2$ (solid).}
\end{figure}
It is important to note that the value of the dilaton function at
 the origin $\Phi(t,r=0)$ is a free parameter and there are no reasons to
keep it equal to its initial value. Indeed, in our numerical
algorithm we have put only the  condition $\Phi(t=0,r=0)=0$ while
$\Phi(t,r=0)$ has been calculated on each   slice $t>0$ according
to the evolution equations (\ref{5a}, \ref{5b}). This means we have
put free boundary condition for $\Phi(t,r=0)$.

The asymptotic behavior of the solutions regular at infinity is
given by the  series expansion:
\begin{eqnarray}
&&f(t,r)_{r \to \infty}=\pm\left(1-\frac{c}{r}+
O(r^{-2})\right),\nonumber\\ [0.1cm]
 &&\Phi(t,r)_{r \to \infty}=\Phi_{\infty}-\frac{d}{r}+O(r^{-4}),
\end{eqnarray}
where $c,d$ and $\Phi_{\infty}$ are constants. This gives the boundary
conditions at the spatial infinity as follows:
\begin{eqnarray} \label{boundary2}
\nonumber &&f(t,r=\infty)=\pm 1,\quad  f\,'(t,r=\infty)=0,\\[0.1cm]
&&\Phi(t,r=\infty)=\Phi_{\infty}, \quad \Phi\,'(t,r=\infty)=0.
\end{eqnarray}

In order to impose the initial conditions in the Cauchy problem
for the system Eqs.~(\ref{5a}, \ref{5b}), it is enough to set the
initial distribution and its derivatives for the YM function
$f(t=0,r)$ only. Then the dilaton distribution $\Phi(t=0,r)$ can
be obtained from the field equation (\ref{5b}), in a way similar
to the Einstein equations, where the dilaton  plays a role of the
relevant metric function now.

So, we always consider smooth regular initial distributions
$f(t=0,r)=h(r)$, where $h(r)$ provides non-vacuum values for the
YM field function  in some  compact region for $r$ outside the
origin. It is convenient to use two different types of the initial
profiles for $h(r)$:
\begin{equation}
    h(r) = 1-A\,r^2\exp\left[-\sigma(r-R)^2\right]\,, \label{h1}
\end{equation}
which is a Gauss-type ($A$, $\sigma$ and $R$ - constants) initial
profile that connects the same YM vacua $f=+1$, and
\begin{equation}
    h(r) = \frac{1-a\,r^2}{1+a\,r^2},  \label{h2}
\end{equation}
which is a kink-type ($a = const.$), that connects two topologically
distinct YM vacua $f=\pm 1$.

After the initial distribution $h(r)$ is fixed, one can define YM
radial  wave, propagating towards the origin (ingoing wave) as
follows:

\begin{eqnarray} \label{initial}
    f(0,r)=h(r),\quad \dot f(0,r)=f\,'(0,r)=h\,'(r).
\end{eqnarray}
We also put:
\begin{equation}
\dot \Phi(0,r)=0.
\end{equation}
The dilaton function at $t=0$ can be obtained now from the Eq.(\ref{5b}) with the initial data at the origin imposed:
\begin{eqnarray}
    &&-\Phi\,''-\frac{2\Phi\,'}{r}=-\frac{e^\Phi}{r^2} \left[\frac{\left( h^2 - 1 \right)^2} {2r^2} \right], \nonumber \\[0.1cm]
&& \Phi(0,r=0)=\Phi\,'(0,r=0)=0. \label{typ1}
\end{eqnarray}

The symmetric initial conditions that lead to two YM radial waves,
propagating towards (ingoing) and outwards (outgoing) the origin
are determined in a similar way:
\begin{eqnarray}
    &&f(0,r)=h(r),\quad \dot f(0,r)=0, \quad \dot\Phi(0,r)=0,
\nonumber \\
&&-\Phi\,''-\frac{2\Phi\,'}{r} = -\frac{e^\Phi}{r^2} \left[h\,'\,^2+ \frac{\left(h^2-1\right)^2}{2r^2} \right], \nonumber\\
 && \Phi(0,r=0)=\Phi\,'(0,r=0)=0. \label{typ2}
\end{eqnarray}
\begin{figure*}
\includegraphics[width=16.cm]{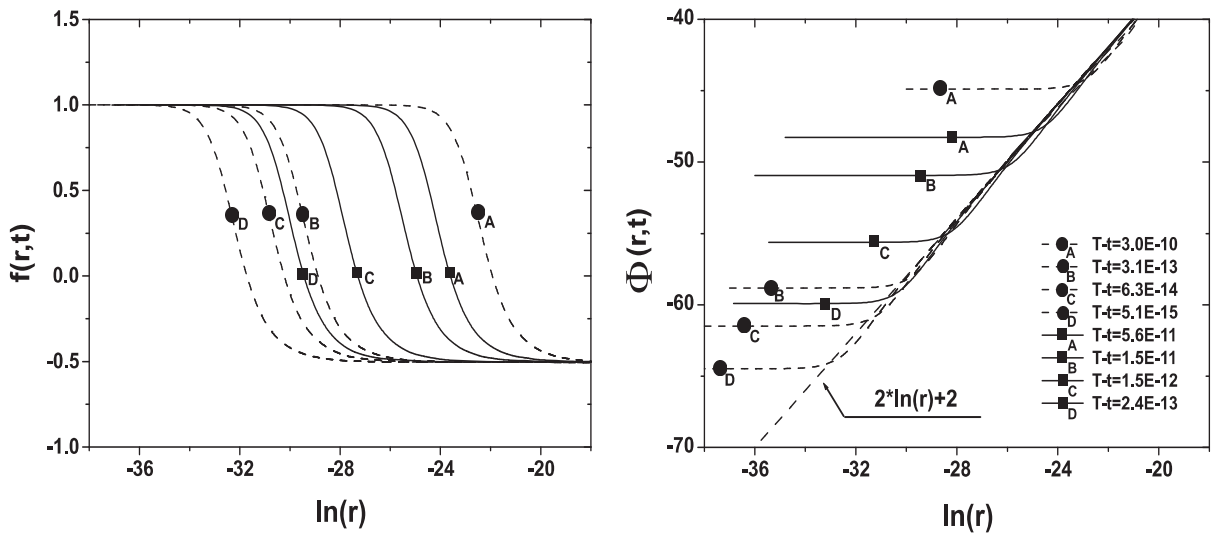}
\caption{\label{fig:4} Self-similar regime of the solutions
behavior: YM function $f(t,r)$ (left), dilaton function $\Phi(t,r)$
(right) at the various $t$ prior to the blowup. The solutions
effectively depends on $T-t$, where $T$ is blowup time.
The initial data is the same as in Fig.\ \ref{fig:3}: kink-type
initial profile (cycles), Gauss-type initial profile (squares).}
\end{figure*}
The Cauchy problem has been studied numerically with the help of a
finite difference scheme that preserves the total energy during
the evolution. Adaptive mesh refinement algorithm was also
installed in order to get  a better resolution on a small scales
prior to the singularity formation.

\begin{figure*}
\includegraphics[width=17.cm]{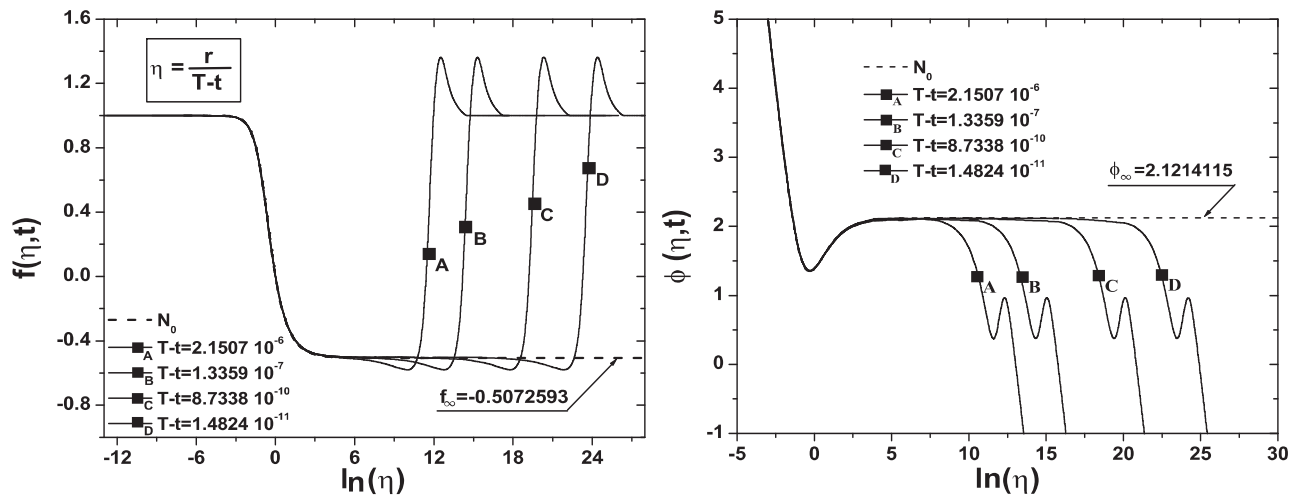}
\caption{\label{fig:5} The late-time evolution of a some typical
Cauchy problem solution prior to the blowup. YM profiles
$f(\eta,t)$ (left) and the scale-invariant part of the dilaton
profiles $\phi(\eta,t)=\Phi(r,t)-2\ln(r)$ --- (right) attain the
self-similar solution $N=0$ (dashed line). The values $f_{\infty},
\phi_{\infty}$ are the asymptotical values of the self-similar
$N=0$ solution at $\eta \to \infty$. The initial conditions are
the same as in Fig.\ \ref{fig:3} for Gauss-type initial profile.}
\end{figure*}

The system has been solved for various initial profiles $h(r)$
and initial data (\ref{typ1}-- \ref{typ2}). The results can
be summarized as follows.
If we start
with  small initial data,  controlled by the values of the
parameters in Eqs.~(\ref{h1}), (\ref{h2}), the ingoing wave
solution get smoothly bounced near the origin and after which is
radiated away qualitatively similarly to the typical solution of a
linear wave equation. If we gradually change the relevant
parameters in the initial conditions towards more strength initial
data, after it exceeds some threshold value,
the ingoing YM wave solution demonstrates universal behavior
at the small scales. In fact, the YM function $f(t,r)$ attaines
the self-similar $N=0$ solution  $f_0(\frac{r}{T-t})$, see
Figs.~\ref{fig:3},~\ref{fig:4}. As it was expected, in this regime
the corresponding solution for the dilaton function $\Phi(t,r)$
develops in a non self-similar way. However, as it can be easily
seen from Fig.~\ref{fig:4}, the whole $\Phi(t,r)$ profile
propagates along the straight line $2\ln{r} + 2$. This means that
in the appropriately shifted and rotated at angle $\beta={\rm
arctg}(-2)$ back clockwise frame the obtained dilaton function
$\phi(t,r)$, $\phi(t,r)=\Phi(t,r) \cos{\beta} +
\sin{\beta}\ln{(r/r_0)},\, r_0=2,$ exhibits the self-similar
behavior and attains the self-similar $N=0$ solution
$\phi_0(t,r)$. The dilaton function $\Phi(t,r)$  can be expressed
now in terms of its self-similar part $\Phi(t,r)=\phi(t,r) + 2\ln
r$ that reproduces the ansatz (\ref{prphi}). Fig.~\ref{fig:5}
illustrates the late-time evolution of some typical solution to
the Cauchy problem which attains the self-similar solution
$f_0(\eta), \phi_0(\eta)$.

 The solutions evolve in a self-similar regime during a
finite time and at some blowup time $T$ the second derivative of
the YM function $f$ exhibits unboundedly growth at the origin.
This is an indication for a singularity formation. Blowup time $T$
is just the total time in the Cauchy problem and it depends on the
initial data. However, as the solutions attain the self-similar
profile $f_0(\eta), \phi_0(\eta)$ their dependence on the time
enters effectively only in the form $T-t$. Hence, the late-time
asymptotics becomes universal for an arbitrary solution with the
initial data, which leads to the blowup. According to our studies,
the self-similar solution $f_0(\eta), \phi_0(\eta)$ is linearly
stable and, as a result, it can pretend to be a global stable
attractor in the Cauchy problem.
The singularity formation at the origin is not accompanied by an
energy concentration which is typical for the supercritical
systems. Indeed, the total energy Eq.~(\ref{Etot1}) at time $t
\leq T$ inside the past light cone $\int_0^{T-t} dr \left(\ldots \right)$ of the point $(T,0)$ can be calculated  using  the $N=0$
self-similar solution $f_{0}(x)$, $\phi_{0}(x)$ as follows:
\begin{widetext}
    \begin{eqnarray}
\nonumber
        M = E =(T-t) \int\limits_1^{\infty}\,dx \left\{
 \frac{x^2+1}{2x^2} \phi_0(x)^2_{,x} -
\frac{2}{x} \phi_0(x)_{,x} + \frac{2}{x^2}+
        e^{\phi_0(x)}\left[\frac{x^2+1}{x^2}  f_0(x)^2_{,x} +
\frac{\left(f_0(x)^2-1\right)^2} {2x^2} \right] \right\} \sim 4(T-t),
    \end{eqnarray}
\end{widetext}
\noindent
and vanishes as $t \to T$. Note that numerically the integral is
equal to 4 with accuracy $10^{-9}$. The corresponding
Schwarzschildean radius is equal to $r_S \sim 8(T-t)$. It means that if
the gravity is included, the obtained late-time self-similar
attractor (if exists) has to be hidden  under the formed event
horizon.

\section{ Conclusions and discussion}

We conclude with the following summary. Using the special ansatz
for the dilaton field, we brought the system of the SU(2)
spherically symmetric Yang-Mills-dilaton equations to a
scale-invariant form and found an infinite countable family of
self-similar solutions, labeled by the $N \geq 0$ -- number of
zeros of the relevant YM function. Among them the only lowest
solution, which corresponds to the $N=0$ is stable in the
framework of the linear perturbation theory. Being a
scale-invariant, the considered system has a criticality index
equal to $+1$ which means that the system is a supercritical one
in the PDE terminology and should exhibit a singularity formation
if the initial data in the evolutional Cauchy problem exceeds some
threshold value. The Cauchy problem has been solved numerically
for a wide range of smooth finite energy initial data and the
results we found are in agreement with this general expectation.
It has been found that if the initial data exceeds some threshold,
the resulting solutions in a compact region, shrinking to the
origin, attain the lowest $N=0$ stable self-similar profile, which
plays the role of an attractor in the Cauchy problem. If the
solutions attain the $N=0$ self-similar attractor they evolve in
an universal way during a finite time until the second derivative
of the YM function at the origin starts growing infinitely.

The problem of the threshold, that separates the disperse and
blowup behavior of the Cauchy problem solutions requires more
detailed studies. This threshold is usually related to the
intermediate attractors, which are some local minima or saddle
points of the effective action. In our system both static $N=1$
\cite{r10} and self-similar solution $f_1(\eta)$, $\phi_1(\eta)$
are saddle points of the effective action and can play the role of
such intermediate attractors. The study of the decay of the static
and the self-similar solutions with $N \geq 1$ along their
unstable modes are of particular interest for the threshold
understanding.

As it was noted earlier, the system of Yang-Mills-dilaton fields
is the bosonic part of the appropriately truncated supersymmetric
field theory which also contains  spin $1/2$ dilatino and gaugino
fields in the fermionic sector. The study of the fermionic fields
in the obtained self-similar bosonic backgrounds seems to be a
very interesting task itself, and also can shed  new light on the
singularity formation thresholds of the bosonic fields.

All these tasks are under considerations and will be reported elsewhere.

\begin{acknowledgments}
We would like to thank E. A. Ayrjan for his kind attention to this
work and helpful discussions. Discussions with A. Barvinski, A.
Gladyshev, D. Fursaev, D. I. Kazakov, A. Pashnev, S. Solodukhin,
M. Tsulaia and E. P. Zhidkov are greatly acknowledged. The work of
T. L. B. was supported in part by Bulgarian Scientific Fund.
\end{acknowledgments}

\end{document}